\documentclass[12pt]{article}
\usepackage{euscript}
\usepackage{amssymb}
\usepackage{amsmath}
\usepackage{graphicx}
\usepackage[small,sc,center]{titlesec}
\usepackage{indentfirst}
\usepackage{siunitx}

\newcommand{\nc}{\newcommand}
\nc{\alphal}{\alpha_\Lambda}
\nc{\alphar}{\alpha_\mathrm{R}}
\nc{\cp}{C_\mathrm{P}}
\nc{\drad}{D_\mathrm{rad}}
\nc{\ek}{E_\mathrm{K}}
\nc{\et}{E_\mathrm{t}}
\nc{\gammar}{\gamma_\mathrm{R}}
\nc{\hp}{H_\mathrm{P}}
\nc{\mzams}{M_\mathrm{ZAMS}}
\nc{\tev}{t_\mathrm{ev}}

\begin{document}

\begin{center}
\textbf{Theoretical Rates of Pulsation Period Change in the Galactic Cepheids}

\textbf{Yu. A. Fadeyev\footnote{E--mail: fadeyev@inasan.ru}}

\textit{Institute of Astronomy, Russian Academy of Sciences, Pyatnitskaya ul. 48, Moscow, 119017 Russia}

Received December 17, 2013

\end{center}

\textbf{Abstract} ---
Theoretical estimates of the rates of radial pulsation period change in Galactic Cepheids
with initial masses $5.5M_\odot\le\mzams\le 13M_\odot$, chemical composition
$X=0.7$, $Z=0.02$ and periods $1.5~\mathrm{day}\le\Pi\le 100~\mathrm{day}$
are obtained from consistent stellar evolution and nonlinear stellar pulsation
computations.
Pulsational instability was investigated for three crossings of
the instability strip by the evolutionary track in the HR diagram.
The first crossing occurs at the post--main sequence helium core gravitational
contraction stage which proceeds in the Kelvin--Helmholtz timescale whereas
the second and the third crossings take place at the evolutionary stage of thermonuclear
core helium burning.
During each crossing of the instability strip the period of radial pulsations
is a quadratic function of the stellar evolution time.
Theoretical rates of the pulsation period change agree with observations
but the scatter of observational estimates of $\dot\Pi$ noticeably exceeds
the width of the band ($\delta\log\vert\dot\Pi\vert\le 0.6$) confining
evolutionary tracks in the period -- period change rate diagram.
One of the causes of the large scatter with very high values of $\dot\Pi$
in Cepheids with increasing periods might be the stars that cross the instability
strip for the first time.
Their fraction ranges from 2\% for $\mzams=5.5M_\odot$ to
9\% for $\mzams=13M_\odot$ and variables $\alpha$~UMi and IX~Cas seem to belong
to such objects.

Keywords: \textit{stars: variable and peculiar.}

\newpage
\section*{introduction}

One of the most remarkable features of $\delta$ Cephei pulsating variable stars
is the exact repetition of the light curve, so that periods of many variables
of this type are known with as many as eight or nine siginificant digits.
At the same time the $O-C$ diagrams of Cepheids reveal the presence of the
quadratic term indicating the secular period change.
Some Cepheids show the secular period decrease, whereas periods of other Cepheids
increase.
The period change rate is highest (in absolute value) in long--period Cepheids
and in the star SV~Vul ($\Pi=45.01$~day) is $\dot\Pi = -214~\mbox{s}/\mbox{yr}$
(Turner, Berdnikov, 2004).

The first report on discover of Cepheid secular pulsation period change
was made by Hertzsprung (1919) who analyzed photometric observations of the
variable star $\delta$~Cephei in the years from 1785 to 1911.
Later the secular pulsation period changes were found in other Cepheids and
the comprehensive review of the studies done in the first half of the XX century
was presented by Parenago (1956).
The interest in Cepheid period changes grew after works by
Hofmeister et al. (1964) and Iben (1966) who established the evolutionary state
of these pulsating  stars and the secular period change of radial oscillations was
explained as a consequence of stellar evolution during thermonuclear core helium
burning.
Moreover, using the approximate estimates Hofmeister (1967) showed that
the rates of the evolutionary change of pulsation periods in Cepheids with
masses $5M_\odot\le M\le 9M_\odot$ are sufficiently large in order to be detected
from available photometric measurements.
By now the observational estimates of $\dot\Pi$ are obtained for about two
hundred Galactic Cepheids (Turner 1998; Turner et al. 2006).

Secular period changes of Cepheids are of great interest since comparison of
observed period change rates with calculations of stellar evolution and radial
stellar pulsation may provide a test of the theory.
Moreover, this phenomenon allows us to roughly evaluate the stellar mass
without the period--mean density relation.
Unfortunately, the theoretical studies of secular Cepheid period changes
have not been done till recent time,
so that comparison between observations and the theory was restricted by
rough approximations (Pietrukowicz 2001; Turner at al. 2006).

The first theoretical rates $\dot\Pi$ obtained from consistent stellar evolution
and nonlinear stellar pulsation calculations were presented in our
previous paper (Fadeyev 2013b) for Cepheid models with fractional mass
abundances of hydrogen and elements heavier than helium $X=0.7$ and $Z=0.008$,
respectively, which are typical for stars of the Large Magellanic Cloud (LMC).
Results of calculations done for LMC Cepheids at the thermonuclear core helium burning
were found to be in a good agreement with observations because
in the period -- period change rate diagrams the theoretical tracks locate
within the domain of observational estimates of $\dot\Pi$.
In the present study we employ the same approach for theoretical estimates of
the period change rates in Galactic Cepheids with initial chemical composition
$X=0.7$, $Z=0.02$.
It should be noted that below we present the theoretical rates $\dot\Pi$
not only for the evolutionary stage of thermonuclear core helium burning when
the evolutionary track loops in the Hertzsprung--Russel (HR) diagram but also
for the shorter post--main sequence evolutionary stage of gravitational
contraction of the helium core which proceeds in the Kelvin--Helmholtz timescale.

\section*{the method of computations}

Investigation of nonlinear stellar pulsations is the solution of the Cauchy problem
for equations of radiation hydrodynamics describing spherically--symmetric motions
of the self--gravitationg gas.
Initial conditions corresponding to the hydrostatic and thermal equilibrium are
determined for the main physical variables (radius, luminosity, pressure, temperature,
element abundances) as a function of the Lagrangean mass coordinate.
To this end we use one of the stellar models of the evolutionary sequence
computed from the zero age main sequence using the method developed by Henyey et al. (1964).
Basic assumptions and details of implementation of this method are given
in one of our previous papers (Fadeyev 2013a).

Convective heat transfer and convective mixing of stellar matter are treated in
the framework of the convection model by B\"ohm--Vitense (1958) with the ratio of
mixing length to pressure scale height $\alphal = \Lambda/\hp = 1.6$.
The radius of the outer boundary of the convective core was assumed to increase
due to convective overshooting by $0.1\hp$.
The diffusion coefficient in semiconvection regions was calculated according to
Langer et al. (1983).
Calculation of the thermonuclear energy generation rate as well as
solution of the equations of nucleosynthesis were done with NACRE reaction rates
(Angulo et al. 1999).

As in our previous paper (Fadeyev 2013b) devoted to LMC Cepheids
the equations of radiation hydrodynamics were solved with taking into account
the effects of turbulent convection treated according to Kuhfu\ss\ (1986).
In contrast to the local steady--state convection model by B\"ohm--Vitense (1958)
which is employed in the stellar evolution calculations
the turbulent convection model by Kuhfu\ss\ (1986) contains additional parameters
in relations for transfer of enthalpy and kinetic energy of turbulent elements,
dissipation of turbulent kinetic energy, exchange of momentum and energy between
gas flow and convective elements due to turbulent viscosity.
Detailed discussion of these parameters is presented by Kuhfu\ss (1986) and
Wuchterl and Feuchtinger (1998).

Among parameters of the Kuhfu\ss\ (1986) convection model one has to primarily
note the parameter $\alpha_\mu$ in the expression for the kinetic turbulent viscosity
\begin{equation}
\label{turb_visc}
\mu = \alpha_\mu \rho \Lambda \et^{1/2} ,
\end{equation}
where $\rho$ is the gas density and $\et$ is the mean specific turbulent
kinetic energy.
Prevailing point of view on choice of the value of $\alpha_\mu$ does not exist.
Olivier and Wood (2005) computed nonlinear oscillations of red supergiants
and showed that the most appropriate value is $\alpha_\mu = 0.5$ since for
$\alpha_\mu = 1$ pulsations decay.
In our previous studies devoted to nonlinear oscillations of yellow
hypergiants, red supergiants and LMC Cepheids the hydrodynamic
computations were carried out with $\alpha_\mu = 0.5$ and in all
cases we obtained agreement with observations (Fadeyev 2011, 2012, 2013a, b).
However in calculations of nonlinear pulsations of Galactic Cepheids the
parameter $\alpha_\mu=0.5$ leads to decay of oscillations within the whole empirical
instability strip and therefore is in conflict with observations.
To clarify the role of this parameter we performed numerical experimets and found
that agreement with observations can be obtained for $0.1\le\alpha_\mu\le 0.25$,
the lower limit of this range corresponding to Cepheids of smaller masses.
It should be noted that nonlinear oscillations of Galactic Cepheids
computed by Koll\'ath et al. (2002), Szab\'o et al. (2007), Smolec and Moskalik (2008),
Baranowski et al. (2009) were computed with $\alpha_\mu$ values of this range.

The role of radiative cooling of convective elements in damping of stellar
oscillations increases with decreasing mass of the Cepheid.
The rate of radiative energy exchange is given by (Wuchterl and Feuchtinger 1998)
\begin{equation}
\drad = 4\sigma \left(\frac{\alphar}{\alphal}\gammar\right)^2
        \frac{g^2 T^3 \et}{\cp \kappa P^2} ,
\end{equation}
where $g$ is gravitational acceleration, $\cp$ is specific heat at constant pressure,
$P$ is total pressure, $T$ is temperature, $\kappa$ is the Rosseland mean opacity,
$\gammar = 2\sqrt{3}$.
In the present study the value of the parameter $\alphar$ was determined from
computational experiments.
For models of massive Cepheids ($\mzams\ge 9M_\odot$) the role of radiative cooling
was found to be negligible and computations were carried out with $\alphar = 1$.
However for Cepheids with masses $\mzams\le 8M_\odot$
the limit cycle oscillations can be obtained only for smaller values of this parameter:
$0.1\le\alphar\le 0.5$.
It should be noted that in their computations 
Koll\'ath et al. (2002), Szab\'o et al. (2007), Smolec and Moskalik (2008),
Baranowski et al. (2009) also used small values of this parameter:
$0\le \alphar < 0.5$.

Parameters $\alpha_\mu$ and $\alphar$ substantially affect the pulsational
instability of Galactic Cepheids but the role of uncertainties in their values
for determination of radial pulsation periods is negligible.
This is due to the fact that excitation and damping of pulsational instability
occur in outer layers of the Cepheid whereas the pulsation period $\Pi$ is
proportional to the sound travel time between the center and the surface of the star.
For example, for Cepheids with initial mass $\mzams=7M_\odot$
at the evolutionary stage of the second crossing of the instability strip
the results of calculations agree with observations for $\alpha_\mu = 0.1$, $\alphar=0.1$
whereas for $\alpha_\mu = 0.5$, $\alphar=1$ oscillations decay.
At the same time the pulsation periods of these two sequences of hydrodynamical
models differ less than one tenth of a percent, that is are the same within the error of
period evaluation.

\section*{results of computations}

\subsection*{evolutionary and hydrodynamical models}

The role of small initial perturbations at $t=0$ in the Cauchy problem for
equations of hydrodynamics is played by interpolation errors arising due to conversion
of the evolutionary stellar model to the Lagrangean grid of the
hydrodynamical model.
Evolutionary stellar models at the Cepheid stage were computed with $\sim 10^4$
mass zones, whereas hydrodynamical computations were carried out with
the number of Lagrangean zones $N=500$.
Lagrangean intervals of the hydrodynamical model increase geometrically inward
from the outer boundary.

The need to compute evolutionary models with the large number of mass zones
is due to the two following reasons.
First, at the core helium burning stage the outer boundary of the convective core
coincides with abundancy jumps of helium, carbon and oxygen.
Therefore the fine spatial grid in the vicinity of the jump of the
mean molecular weight allows us to diminish the amplitude of sharp changes of the central
energy generation rate that accompany the mass growth of the convective core
due to discrete representation of the evolutionary model.
Second, the large number of mass zones in outer layers of the evolutionary
model leads to smaller amplitude of initial perturbations that are
proportional to interpolation errors when the evolutionary model is converted to
the hydrodynamical model.

At each step of integration of the equations of hydrodynamics with respect to time $t$
we computed the kinetic energy $\ek(t)$ of the gas flow in the stellar envelope.
Within initial time interval with exponential change of $\ek(t)$ we evaluated
the instability growth rate $\eta = \Pi d\ln\ek/dt$.
To determine the period of radial oscillations $\Pi$ we applied the
discrete Fourier transform of the kinetic energy $\ek(t)$ within the time interval
involving more than $10^2$ pulsation cycles.

In the present study we computed nine evolutionary tracks for stars with initial
masses $5.5M_\odot\le\mzams\le 13M_\odot$.
For initial conditions of the Cauchy problem of the equations of hydrodynamics we
used several dozen evolutionary models with luminosities and effective temperatures
close to the empirical instability strip of Cepheids in the HR diagram.
The method of determination of the boundaries of the theoretical pulsation
instability strip is described in our previous paper (Fadeyev 2013b).

At the evolutionary stage of thermonuclear helium burning
(i.e. the second and the third crossings of the pulsation instability strip)
the mass $M$ and the luminosity $L$ of Cepheid evolutionary models
approximately obey the relation
\begin{equation}
\label{l-m}
\log (L/L_\odot) = 0.397 + 3.832 \log (M/M_\odot) .
\end{equation}
The mass--luminosity relation obtained by Baranowski et al. (2009)
from evolutionary tracks by Schaller et al. (1992) leads to the luminosity
higher than that from relation (\ref{l-m}) by $0.1\le\delta\log L\le 0.15$.
This difference is mainly due to the fact that Schaller et al. (1992)
computed their evolutionary models with overshooting distance ($0.2\hp$)
which is larger than that in the present study.

\begin{figure}
\centerline{\includegraphics[width=15cm]{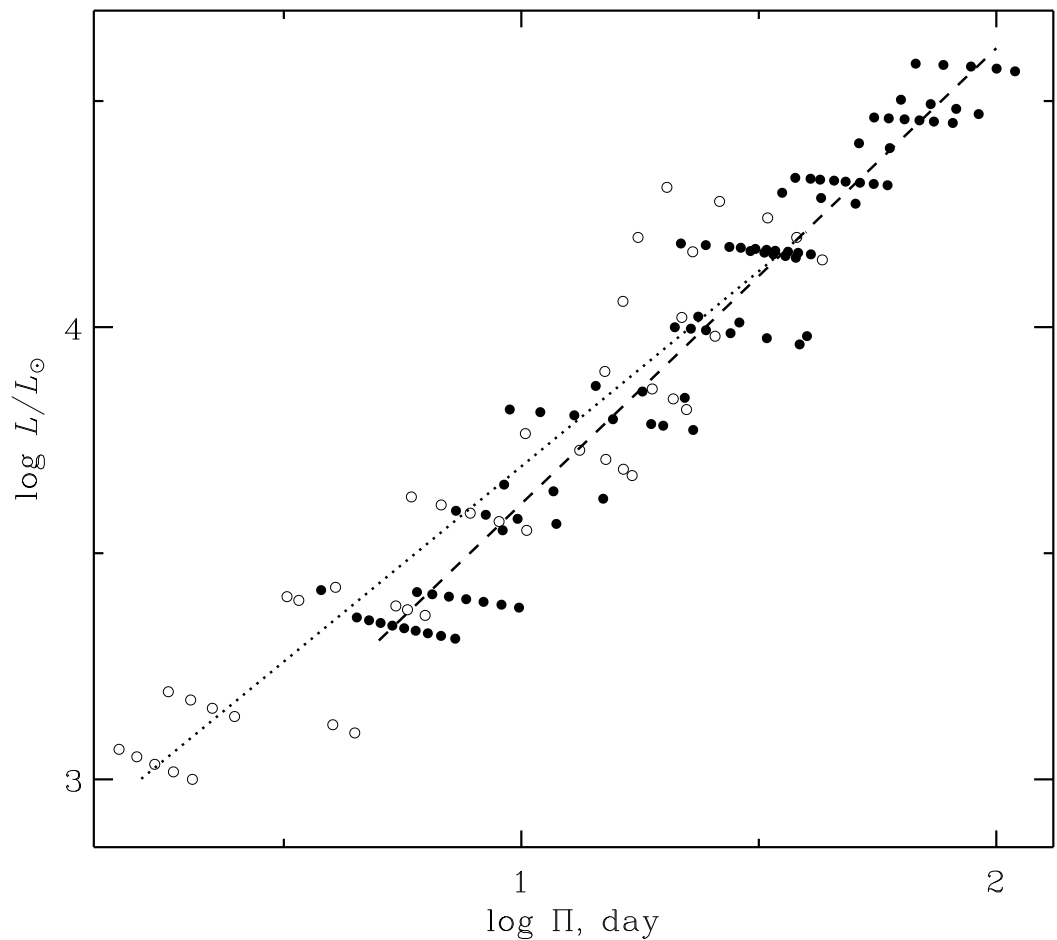}}
\caption{Hydrodynamical models of Galactic Cepheids in the period--luminosity diagram.
         Cepheids of the first crossing of the instability strip are shown in open
         circles and Cepheids of the second and the third crossings are shown in
         filled circles.
         Relations (\ref{lm1}) and (\ref{lm23}) are shown by dotted and dashed lines,
         respectively.
}
\label{fig1}
\end{figure}

All Cepheid hydrodynamical models computed in the present study and that
were found to be unstable against radial oscillations are
displayed in the period--luminosity diagram in Fig.~\ref{fig1}.
As is seen from the plots the Cepheid models can be divided into two groups.
Cepheids of the first group are at the post--main sequence evolutionary stage
of gravitational contraction of the helium core (i.e. the first crossing of the
instability strip).
In Fig.~\ref{fig1} these models are shown in open circles.
Cepheids of the second group are at the evolutionary stage of thermonuclear core
helium burning (i.e. the second and the third crossings of the instability strip)
and in Fig.~\ref{fig1} they are shown in filled circles.
Dependence of the luminosity on the pulsation period for each group of
Cepheids is given by relations
\begin{align}
\label{lm1}
&\log (L/L_\odot) = 2.829 + 0.863 \log\Pi , & (\mbox{crossing 1}) ,
\\
\label{lm23}
&\log (L/L_\odot) = 2.602 + 1.008 \log\Pi , & (\mbox{crossings 2, 3}) ,
\end{align}
where the period $\Pi$ is expressed in days.

\begin{figure}
\centerline{\includegraphics[width=15cm]{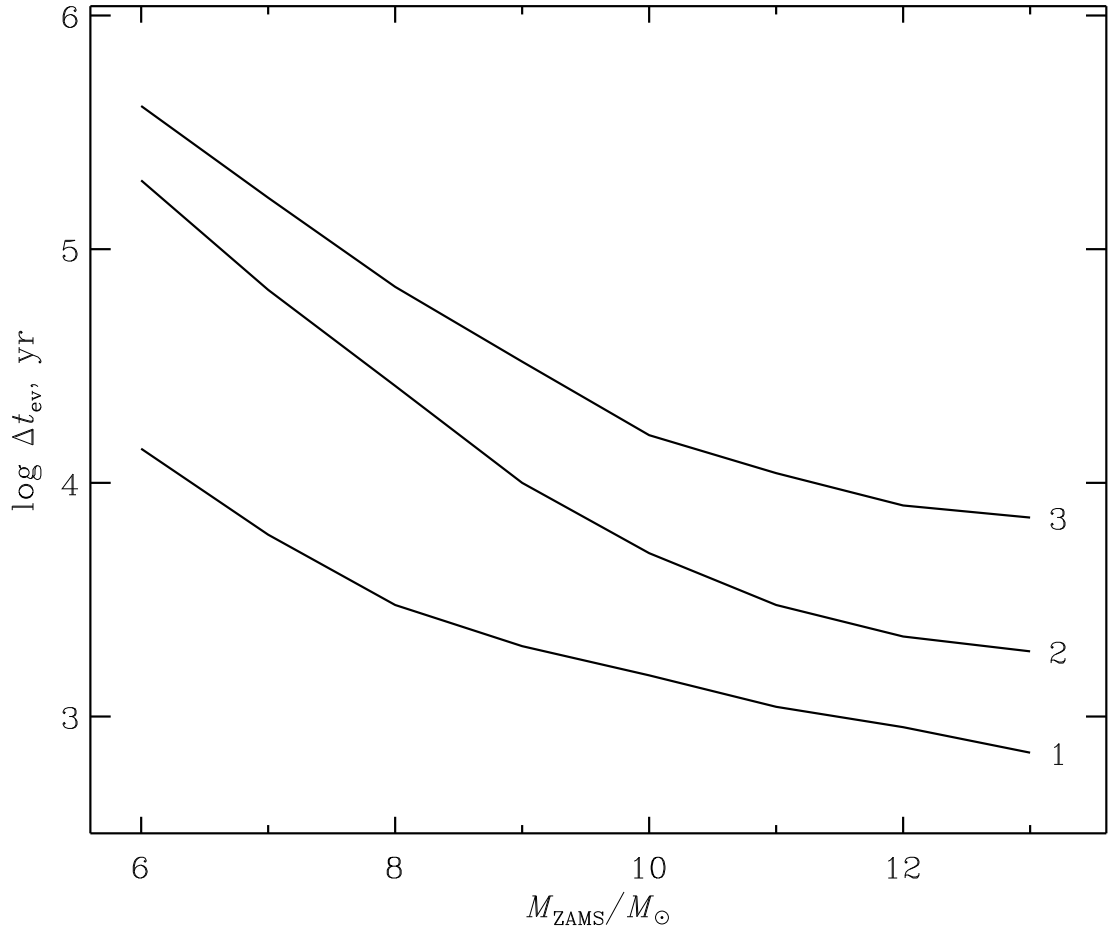}}
\caption{The time spent by the star in the instability strip $\Delta\tev$ as
         a function of the initial stellar mass $\mzams$.
         Digits at the curves indicate the number of the instability strip crossing.
}
\label{fig2}
\end{figure}

Here one should bear in mind that the role of Cepheids of the first crossing
in the period--luminosity diagram is quite small due to the shorter Kelvin--Helmholtz
timescale in comparison with helium burning time.
In order to roughly evaluate the probability to observe the Cepheid with
gravitationally contracting helium core we compare times $\Delta t_{\mathrm{ev},i}$
spent by the star in the instability strip.
The plots of $\Delta t_{\mathrm{ev},i}$ as a function of the initial stellar mass $\mzams$
are shown in Fig.~\ref{fig2}.
Therefore, the fraction of Cepheids of the first crossing among all Cepheids
with initial mass $\mzams$ ranges within
$
0.02\le
\Delta t_\mathrm{ev,1}/\left(\Delta t_\mathrm{ev,1}+\Delta t_\mathrm{ev,2}+\Delta t_\mathrm{ev,3}\right)
\le 0.07
$
for $6M_\odot\le\mzams\le 13M_\odot$,
and this ratio increases with increasing stellar mass.

\subsection*{the change of radial pulsation periods}

In our previous paper (Fadeyev 2013b) devoted to LMC Cepheids we showed that
during evolution across the instability strip the period of radial oscillations
is described by a quadratic function of the evolution time $\tev$ with
relative error less than one per cent.
Hydrodynamical calculations of the present study showed that this conclusion
is also valid for Galactic Cepheids at all three crossings of the instability
strip.
This allows us to easily obtain reliable estimates of the period change rate
$\dot\Pi$ as a function of the evolution time.

Let us consider the results of computations in the period--period change rate
diagram where each sequence of hydrodynamical models with $\eta > 0$
is displayed by the curve.
Theoretical dependences for Cepheids of the second crossing of the instability strip
($\dot\Pi < 0$) are shown in Fig.~\ref{fig3}, whereas in Fig.~\ref{fig4} we
give the dependences for Cepheids of the first and of the third crossings ($\dot\Pi > 0$)
that are shown in dashed and solid lines, respectively.
In these diagrams we plot also the observational estimates of $\Pi$ and $\dot\Pi$ for
50 Cepheids with decreasing periods and 93 Cepheids with increasing periods.
Observational estimates are taken from the paper by Turner (1998) with
tabular data for $\Pi$ and $\dot\Pi$ of 137 Galactic Cepheids.
More recent observational estimates of $\dot\Pi$ were taken from papers by
Berdnikov and Turner (2004), Turner and Berdnikov (2004),
Berdnikov et al. (2006), Berdnikov and Pastukhova (2012).

In the period--period change rate diagram
the evolutionary tracks with $\eta > 0$ locate along the straight lines given by
\begin{equation}
\label{pi-dotpi}
\log\vert\dot\Pi\vert =
\left\{
\begin{array}{l}
-7.57 + 2.15 \log\Pi ,
\\
-9.19 + 2.69 \log\Pi ,
\\
-9.92 + 2.79 \log\Pi
\end{array}
\right.
\end{equation}
for the first, the second and the third crossings of the instability strip,
respectively.
Here the period change rate $\dot\Pi$ is dimensionless and the radial pulsation
period $\Pi$ is expressed in days.
The width of the band which confines the tracks in the
($\log\Pi,\log\vert\dot\Pi\vert$)
plane is $\delta\log\vert\dot\Pi\vert = 0.6$ for the first and for the second
crossings of the instability strip, whereas for the third crossing
$\delta\log\vert\dot\Pi\vert = 0.4$.
The fraction of Cepheids with increasing periods of the first crossing
of the instability strip is given by the ratio
$\Delta t_\mathrm{ev,1}/(\Delta t_\mathrm{ev,1} + \Delta t_\mathrm{ev,3})$
and increases from 2\% for Cepheids with initial mass $\mzams=5.5M_\odot$
to 9\% for $\mzams=13M_\odot$.

Pulsation periods $\Pi$ and pulsation period change rates $\dot\Pi$ for Cepheids
with intial masses $6M_\odot\le\mzams\le 12M_\odot$ are given in the table.
Each crossing of the instability strip in the HR diagram is represented by two
pairs of the values of $\Pi$ and $\dot\Pi$.
The first pair corresponds to the initial point of the track in the
($\Pi,\dot\Pi$) plane and the second one corresponds to the final pont
of the track.
For the sake of more convenient comparison of the tabular data with
observations the pulsation period $\Pi$ is expressed in days and the
period change rate $\dot\Pi$ is given in units of seconds per year.

\subsection*{conclusions}

The period--period change rate diagrams in Fig.~\ref{fig3} and Fig.~\ref{fig4}
show that theoretical estimates of the pulsation period change rates
of Galactic Cepheids are in a good agreement with observations
and this conclusion is an important argument in favour of the theory of stellar
evolution.
At the same time one has to emphasize the fact that the scatter of
observational estimates of $\dot\Pi$ substantially exceeds the width of
the band confining theoretical tracks.
The same property also show LMC Cepheids (Fadeyev 2013b).
One of the causes of the large scatter of observational estimimates of $\dot\Pi$
seem to be errors which increase with decreasing period because of the power
relationship (\ref{pi-dotpi}) between the pulsation period $\Pi$ and the period
change rate $\dot\Pi$.

Another cause of the large scatter of observational estimates of the
period change rate for $\dot\Pi > 0$ might be Cepheids undergoing
the post--main sequence gravitational contraction of the helium core.
The number of such stars among Cepheids with increasing periods is roughly
several per cents.
Pulsating variable stars
$\alpha$~UMi ($\Pi=3.97$~day, $\dot\Pi = 3.2$~s/yr)
and IX~Cas ($\Pi=9.16$~day, $\dot\Pi = 64.27$~s/yr)
seem to belong to such objects.

The study was supported by the Basic Research Program of the Russian Academy of Sciences
``Nonstationary phenomena in the Universe''.

\newpage
\subsection*{REFERENCES}
\begin{enumerate}

\item C. Angulo, M. Arnould, M. Rayet, et al., Nucl. Phys. A \textbf{656}, 3 (1999).

\item R. Baranowski, R. Smolec, W. Dimitrov, et al., MNRAS \textbf{396}, 2194 (2009).

\item L.N. Berdnikov and D.G. Turner, Astron. Astrophys. Trans. \textbf{23}, 123 (2004).

\item L.N. Berdnikov, E.N. Pastukhova, N.A. Gorynya, et al., Astron. Astrophys. Trans. \textbf{25}, 221 (2006).

\item L.N. Berdnikov and E.N. Pastukhova, Astron. Zh. \textbf{89}, 931 (2012)
      [Astron. Rep. \textbf{56}, 843 (2012)].

\item E. B\"ohm--Vitense, Zeitschrift f\"ur Astrophys. \textbf{46}, 108 (1958).

\item Yu.A. Fadeyev, Pis'ma Astron. Zh \textbf{37}, 440 (2011)
      [Astron. Lett. \textbf{37}, 403 (2011)].

\item Yu.A. Fadeyev, Pis'ma Astron. Zh \textbf{38}, 295 (2012)
      [Astron. Lett. \textbf{38}, 260 (2012)].

\item Yu.A. Fadeyev, Pis'ma Astron. Zh. \textbf{39}, 342 (2013a)
      [Astron.Lett. \textbf{39}, 306 (2013a)].

\item Yu.A. Fadeyev, Pis'ma Astron. Zh. \textbf{39}, 829 (2013b)
      [Astron.Lett. \textbf{39}, 746 (2013b)].

\item L.G. Henyey, J.E. Forbes, and N.L. Gould, Astrophys. J. \textbf{139}, 306 (1964).

\item E. Hertzsprung, Astron. Nachr. \textbf{210}, 17 (1919).

\item E. Hofmeister, Zeitschrift f\"ur Astrophys. \textbf{65}, 164 (1967).

\item E. Hofmeister, R. Kippenhahn, and A. Weigert, Zeitschrift f\"ur Astrophys. \textbf{60}, 57 (1964).

\item I. Iben, Astrophys.J. \textbf{143}, 483 (1966).

\item Z. Koll\'ath, J.R. Buchler, R. Sz\'abo, et al., Astron. Astrophys. \textbf{385}, 932 (2002).

\item R. Kuhfu\ss, 1986, Astron. Astrophys. \textbf{160}, 116 (1986).

\item N. Langer, K.J. Fricke, and D. Sugimoto, Astron. Astrophys. \textbf{126}, 207 (1983).

\item E.A. Olivier and P.R. Wood, MNRAS \textbf{362} 1396 (2005).

\item P.P. Parenago, Perem. Zv. \textbf{11}, 236 (1956).

\item P. Pietrukowicz, Acta Astron. \textbf{51}, 247 (2001).

\item G. Schaller, D. Schaerer, G. Meynet, et al., Astron. Astrophys. Suppl. Ser. \textbf{96}, 269 (1992).

\item R. Smolec, P. Moskalik, Acta Astron. \textbf{58}, 193 (2008).

\item R. Szab\'o, J.R. Buchler, and J. Bartee, Astrophys. J., 667, 1150 (2007).

\item D.G. Turner, JAAVSO \textbf{26}, 101 (1998).

\item D.G. Turner and L.N. Berdnikov, Astron. Astrophys. \textbf{423}, 335 (2004).

\item D. Turner, M. Abdel--Sabour Abdel--Latif, and L.N. Berdnikov, Publ. Astron. Soc. Pacific \textbf{118}, 410 (2006).

\item G. Wuchterl and M. U. Feuchtinger, Astron. Astrophys. \textbf{340}, 419 (1998).
\end{enumerate}

\newpage
The pulsation period $\Pi$ and the period change rate $\dot\Pi$ at the edges of the instability strip

\begin{center}
\begin{tabular}{r|r|S|S|S|S}
\hline
 $\mzams/M_\odot$ & $i$ & ${\Pi,~\mathrm{day}}$  & ${\dot\Pi,~\mathrm{s/yr}}$  & ${\Pi,~\mathrm{day}}$ & ${\dot\Pi,~\mathrm{s/yr}}$ \\
\hline
  6   &  1  &   1.89  &    3.3  &   2.652  &    6.04  \\
      &  2  &   6.82  &    1.6  &   4.436  &    0.487 \\
      &  3  &   6.52  &    0.46 &   9.864  &    1.56  \\[4pt]
  8   &  1  &   6.39  &   46.2  &   8.615  &   82.10  \\
      &  2  &  20.06  &   55.0  &   9.229  &   16.99  \\
      &  3  &  14.28  &    3.9  &  22.515  &   23.64  \\[4pt]
 10   &  1  &  12.56  &  382.6  &  19.229  &  385.32  \\
      &  2  &  48.35  &  460.3  &  26.831  &  283.27  \\
      &  3  &  31.49  &   66.7  &  46.645  &  151.52  \\[4pt]
 12   &  1  &  22.12  &  895.9  &  32.867  & 1167.15  \\
      &  2  &  92.98  & 2165.6  &  51.125  & 1121.45  \\
      &  3  &  49.96  &  113.2  &  72.64   &  539.80  \\
\hline
\end{tabular}
\end{center}
\clearpage

\newpage
\section*{Figure captions}

\begin{figure}
\centerline{\includegraphics[width=15cm]{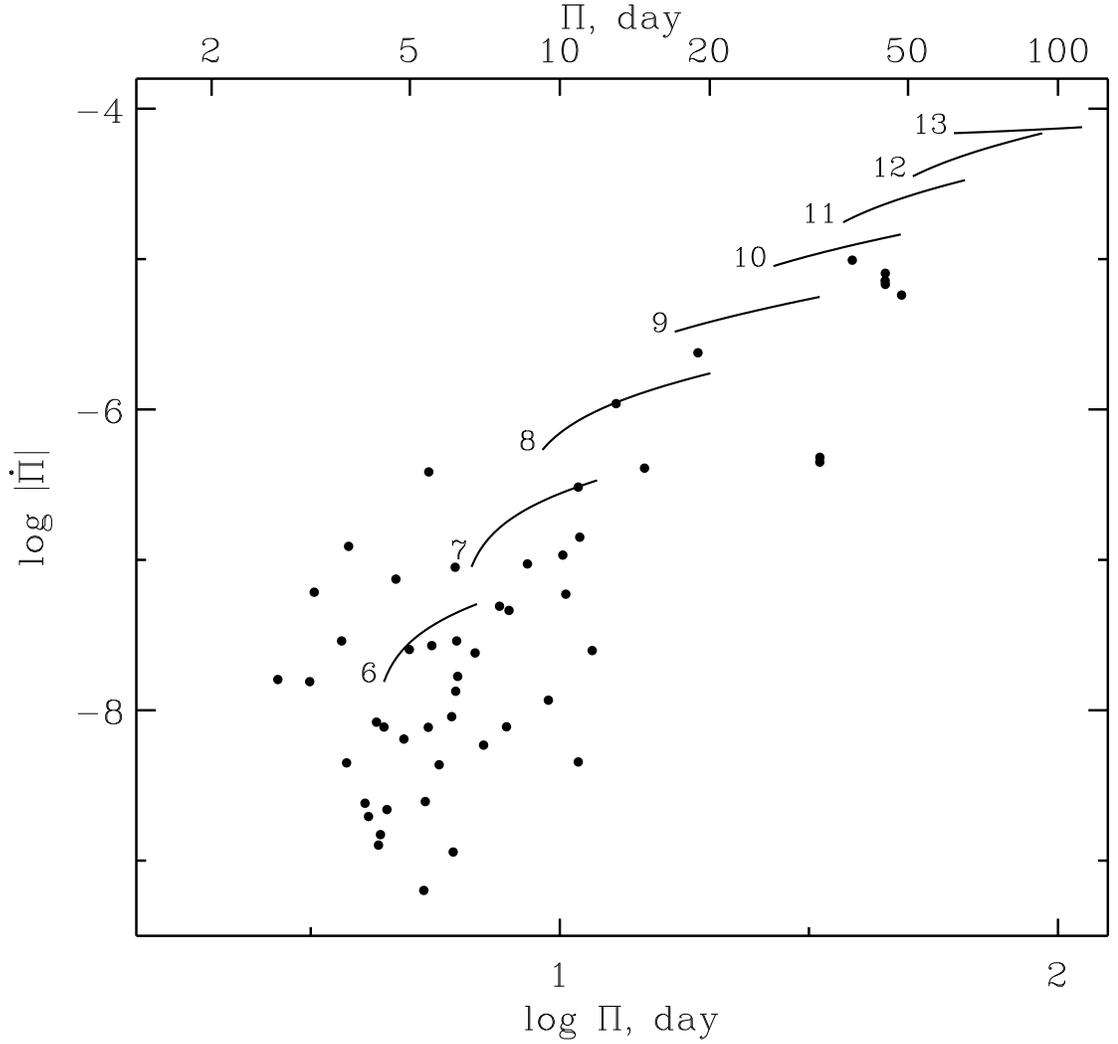}}
\caption{Dimensionless period change rate $\dot\Pi$ as a function of the period
         of radial pulsations $\Pi$ for Galactic Cepheids of the second crossing
         of the instability strip ($\dot\Pi < 0$).
         Parts of evolutionary tracks with $\eta > 0$ are shown in solid lines.
         The value of $\mzams$ is indicated at the tracks.
         Observational estimates of $\Pi$ and $\dot\Pi$ are shown in filled circles.
}
\label{fig3}
\end{figure}

\begin{figure}
\centerline{\includegraphics[width=15cm]{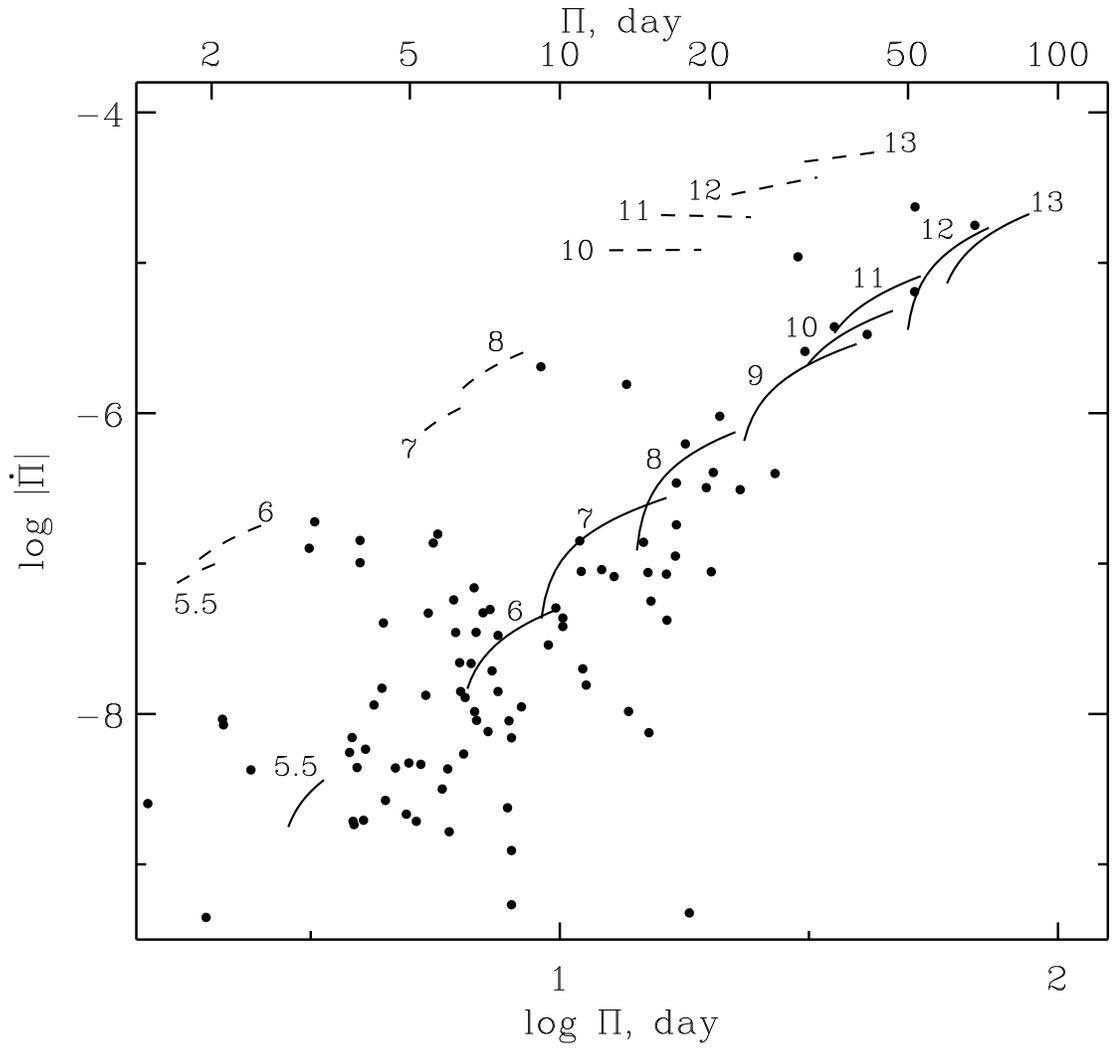}}
\caption{Same as Fig.~\ref{fig3} but for Cepheids of the first (dashed lines)
         and the third (solid lines) crossings of the instability strip ($\dot\Pi > 0$).
}
\label{fig4}
\end{figure}

\begin{itemize}
\item[Fig. 1.]
         Hydrodynamical models of Galactic Cepheids in the period--luminosity diagram.
         Cepheids of the first crossing of the instability strip are shown in open
         circles and Cepheids of the second and the third crossings are shown in
         filled circles.
         Relations (\ref{lm1}) and (\ref{lm23}) are shown by dotted and dashed lines,
         respectively.

\item[Fig. 2.]
         The time spent by the star in the instability strip $\Delta\tev$ as
         a function of the initial stellar mass $\mzams$.
         Digits at the curves indicate the number of the instability strip crossing.

\item[Fig. 3.]
         Dimensionless period change rate $\dot\Pi$ as a function of the period
         of radial pulsations $\Pi$ for Galactic Cepheids of the second crossing
         of the instability strip ($\dot\Pi < 0$).
         Parts of evolutionary tracks with $\eta > 0$ are shown in solid lines.
         The value of $\mzams$ is indicated at the tracks.
         Observational estimates of $\Pi$ and $\dot\Pi$ are shown in filled circles.

\item[Fig. 4.]
         Same as Fig.~\ref{fig3} but for Cepheids of the first (dashed lines)
         and the third (solid lines) crossings of the instability strip ($\dot\Pi > 0$).

\end{itemize}

\newpage

\newpage

\newpage

\newpage

\end{document}